# Room temperature Ferromagnetism in $Th_{1-x}Fe_xO_{2-d}$ (x = 0.0, 0.05, 0.10, 0.15, 0.20 and 0.25) nanoparticles

O. D. Jayakumar[†], I. K. Gopalakrishnan[†]*, A. Vinu[#], A. Asthana[$] and A. K. Tyagi[†]

[†]Chemistry Division, Bhabha Atomic Research Centre, Mumbai-400 085, India.

[#]Nano-Ionics Materials Group, Fuel Cell Materials Center, National Institute for Materials Science, 1-1, Namiki, Tsukuba, Ibaraki 305-0044, Japan.

[$]Advanced Electron Microscopy Group, National Institute for Materials Science, 1-1, Namiki, Tsukuba, Ibaraki 305- 0044, Japan.

## Abstract

Nanocrystalline $(Th_{1-x}Fe_x)O_{2-d}$ particles with different Fe concentrations (x = 0.0, 0.05, 0.10, 0.15, 0.20 and 0.25) have been prepared by a gel combustion method. Rietveld refinement analyses of X-ray diffraction data revealed the formation of an impurity free cubic type $Th_{1-x}Fe_xO_{2-d}$ structure up to x = 0.20. This observation is further confirmed from the detailed studies conducted on 10 at. % Fe doped $ThO_2$ using high-resolution transmission electron microscopy (HRTEM) imaging and indexing of the selected-area electron diffraction (SAED) patterns. *DC* Magnetization studies as a function field indicate that they are ferromagnetic with Curie temperature ($T_c$) well above room temperature.

[*]Corresponding author Email: ikgopal@barc.gov.in

## Introduction

Diluted magnetic semiconductors (DMSs) are the active components of the proposed spintronic devices [1,2]. Since the theoretical prediction of room-temperature ferromagnetism (RTF) in Mn doped p-type ZnO and GaN by Dietl [3] and co-workers, extensive studies have been carried out by many groups on transition metal (TM) doped semiconducting or insulating oxides such as ZnO [4-6], $TiO_2$ [7-9], $SnO_2$ [10], $In_2O_3$ [11,12], $HfO_2$ [13], $CeO_2$ [14] and III-V semiconductor GaN [15,16]. Despite these extensive studies on different types of oxide based DMSs, a clear picture about the origin of ferromagnetism (FM) in these compounds is yet to emerge. However, a number of recent reports by different groups on bulk as well thin films of TM doped semiconducting or insulating oxides converged on to a conclusion that structural defect plays an important role in the origin of FM [5,9-14,17-20]. The doped magnetic ions, mostly 3d transition metals, exhibit a very low solubility in host semiconductor oxides like ZnO and $SnO_2$. This motivated researchers to look for other oxide systems based on host semiconductors with a high solubility of magnetic ions. Yoo et al.[11], have shown that Fe doped $In_2O_3$ co-doped with Cu (Fe up to 20 at. %) is one of the oxide systems that shows promising results as a potential candidate for DMS. Recently, we were able to induce ferromagnetism in Fe doped $In_2O_3$ (Fe concentration up to 20 at. %) without introducing Cu as the additional dopant [21].

Most of the DMS materials developed so far have noncubic crystal symmetry and were either semiconductors or wide band gap semiconductors. Recently Tiwari et al. [14] observed ferromagnetism in Co doped $CeO_2$ films. In this paper, we report the observation of ferromagnetism above room temperature in an insulator $ThO_2$ (band gap =

5.75 eV) doped with high concentration of $Fe^{3+}$ ions (≤ 25 at. %), with face centered cubic fluorite type structure, synthesized by a gel combustion method.

**Experimental Section**

Nanocrystalline particles of $(Th_{1-x}Fe_x)O_{2-d}$ (x = 0.0, 0.05, 0.10, 0.15, 0.20 and 0.25) were prepared by a gel combustion method using glycine as a fuel and nitrates of thorium and iron as the oxidizer. Appropriate amounts of nitrates of Fe and Th were weighed and dissolved in distilled water along with glycine solution (metal to glycine molar ratio 1:1.5). The above solution was evaporated on a hot plate by heating at about 373 K and the gel was obtained. The resultant gel was subsequently swelled into foam undergoing a strong self-propagating combustion reaction to give fine powder. ?-$Fe_2O_3$ powders were also prepared by the same method for comparison. Cationic composition was determined by energy dispersive X-ray analysis. Phase purity and the structure of the fine powder obtained were analyzed by powder X-ray diffraction using CuKα radiation of a Philips Diffractometer (model PW 1071) fitted with graphite crystal monochromater. The lattice parameters of the compounds were extracted by Rietveld refinement of the XRD data by using the computer code Fullprof [22] with the X-ray intensity collected for the range 20° ∠ 2θ ∠ 70°. *DC* magnetization measurements, as a function of field were carried out using an E.G.&G P.A.R vibrating sample magnetometer (model 4500). The electron diffraction and high-resolution transmission electron microscopic (HRTEM) images were acquired on a TEM JEOL 2010F. The preparation of samples for HRTEM analysis involved sonication in ethanol for 2 to 5 min and deposition on a carbon-coated copper grid. The accelerating voltage of the electron beam was 300 kV.

## Results and Discussion

Energy dispersive X-ray analysis (EDAX) indicates that all samples show the nominal concentration of Fe within the experimental error. Rietveld profile refinement analysis of XRD data of $(Th_{1-x}Fe_x)O_2$ (x = 0.0, 0.05, 0.10, 0.15, 0.20 and 0.25) samples revealed the formation of an impurity free cubic fluorite type structure (space group Fm3m) up to x = 0.20 (Fig.1). For concentration above 20 at. % of Fe, peaks corresponding to impurity phase like ?-$Fe_2O_3$ is seen in the XRD patterns (Fig.2). When the same method of combustion is applied to synthesis the oxides of Fe, such as ?-$Fe_2O_3$, the XRD pattern of the sample showed the peaks corresponding to a mixture of α and ?-$Fe_2O_3$ (Fig.2). The crystallite size calculated for pristine $ThO_2$ using Scherrer's equation is approximately 10-15 nm. It is worth to note that the peaks get narrowed down with increase in 'Fe' concentration indicating the increase in crystallinity and particle size as a function of Fe concentration for the samples prepared under identical conditions. This indicates that Fe may perhaps acts as a grain growth promoter on the growth of $ThO_2$ particles and the crystallinity of the resultant product increases. The lattice parameters extracted by Rietveld refinement analysis presented in the inset of Fig.1, shows that the lattice parameter '*a*' remained unchanged with increase of Fe concentration up to *x* = 0.20, even though the ionic radius of $Fe^{3+}$ (0.67 Å) is smaller than that of $Th^{4+}$ ion (1.02 Å). Similar type of behavior is seen by Hanic et al.[23] in CaO-$ThO_2$ system. Using powder X-ray diffraction method and density measurements, they have proved that CaO-$ThO_2$ system contains substitutional as well as interstitial cations. They observed that part of $Ca^{2+}$ enters the substitutional special site (4a): (0, 0, 0) for $Th^{4+}$, while simultaneously, another part of $Ca^{2+}$ ions enters the interstitial positions (4b). Because of this the lattice parameter and hence the unit cell volume remained approximately constant within the concentration range 0 < x < 0.38. Electro neutrality of the solid solution is maintained by the formation of anion vacancies in the oxygen positions (8f,b). We feel it is reasonable to presume that similar explanation holds good for Fe doped $ThO_2$ also.

High-resolution TEM was performed on 10 at. % Fe doped $ThO_2$ particles (taken as a representative sample) to investigate the different phases that might have formed in the nanosize range and to determine the chemical state of Fe atoms which could not be detected by XRD. Figure 3 shows the HRTEM image of 10 at. % Fe doped $ThO_2$ particles. We found no evidence for the presence of Fe metal clusters or Fe-rich secondary phases in these samples, and thus the Fe is homogenously distributed in $ThO_2$ matrix. This is further confirmed from the selected-area electron diffraction (SAED) image showing an impurity free structure (inset top right). Indexing of SAED patterns matched with $ThO_2$ lattice. The particle size of about 10 nm obtained from TEM image matches well with the XRD results.

Figure 4 depicts the room temperature *DC* magnetization measurements of $Th_{1-x}Fe_xO_{2-\delta}$ samples as a function of field. It can be seen from the figure that all samples showed symmetric hysteretic loops typical of ferromagnetic materials. The saturation magnetization value *Ms* is found to increase with Fe concentration up to x = 0.20. However, for compounds with x above 0.20, a decrease in magnetic moment ($\mu_B$/Fe) is observed. It is worth to note that while the large coercive field (H*c* = 750 Oe) remained more or less the same for all compositions of Fe, the remanence increased linearly with Fe content. It is also worth to note that the coercive field is rather small (H*c* = 50 Oe) for $Fe_2O_3$ prepared by the same method (Fig.4). In the present study, we could observe ferromagnetism for Fe concentration right from 5 to 25 % without any co-doping.

Fig.5 depicts the *M vs H* curve for the representative sample 10 at. % Fe doped $ThO_2$, measured at RT and 77 K. It can be seen that the $M_s$ values at 77 K and RT are more or less same. This indicates there is no change in the ferromagnetic exchange interaction at RT and 77 K. This also indirectly indicates that $T_c$ of these compounds is well above RT.

RTF in dilute magnetic oxide systems is a new phenomenon, which challenges our current understanding of magnetic order in solids. Two most ubiquitous models proposed to explain ferromagnetism in DMS viz. carrier induced exchange interaction (RKKY interaction) [25] and double exchange interaction [26] are neither flexible nor accurate enough to describe the properties of real systems. The measured carrier density in real systems is generally not sufficiently high for carrier induced exchange interaction. From the insulator nature of $ThO_2$ (band gap = 5.75 eV) one can rule out carrier induced exchange interaction being the mechanism for the magnetic property observed in $Fe^{3+}$ doped $ThO_2$. The absence of any impurity peaks in the XRD patterns up to 20 at. % of Fe doped $ThO_2$, excludes the possibility of impurity being the source for the observed room temperature ferromagnetism (RTF). Further, the decrease of magnetic moment ($\mu_B$/Fe) observed for Fe concentration above 20 at. % rules out the possible impurities ?-$Fe_2O_3$ and $Fe_3O_4$ being the reason for the observed RTF. Recently Coey et al. proposed F-centre mediated exchange [18,19] to explain the ferromagnetism in diluted magnetic insulators (DMI). An F-centre consists of an electron trapped in an oxygen vacancy. The interaction of magnetic ions and this electron form bound magnetic polarons. As the density of F-centre increases they overlap and produce donor impurity band. In Fe doped $ThO_2$, when $Fe^{3+}$ ions substitute for $Th^{4+}$ ions, oxygen vacancies are formed near the Fe sites to compensate for the imbalance of the charge. $Fe^{3+}$ ions occupying $Th^{4+}$ sites are exchange coupled via electron trapped by charge compensating oxygen vacancies (F-centers) near the $Fe^{3+}$ ions. In other words the oxygen vacancies near $Fe^{3+}$ play a key role in mediating the RTF in the $(Th_{1-x}Fe_x)O_{2-d}$ system. Thus we feel, it is reasonable to speculate that the RTF observed in $Fe^{3+}$ doped $ThO_2$ can be attributed to the oxygen vacancies and /or defects, in agreement with the model proposed by Coey et al.[18,19], and not due to any impurities formed during the synthesis as evidenced by HRTEM, SAED and XRD.

## Conclusion

In conclusion, nanocrystalline particles of $(Th_{1-x}Fe_x)O_{2-\delta}$ (x = 0.0, 0.05, 0.10, 0.15, 0.20 and 0.25) have been synthesized by a gel combustion method. Powder X-ray diffraction studies showed an impurity free mono phasic compound up to x = 0.20. Lattice parameter values are found to remain constant throughout the Fe concentration range. Magnetization studies as a function of field showed ferromagnetism above room temperature in $(Th_{1-x}Fe_x)O_{2-\delta}$ (0.0 ≤ x ≤ 0.25) system, which can be attributed to the oxygen vacancies and/or defects mediated

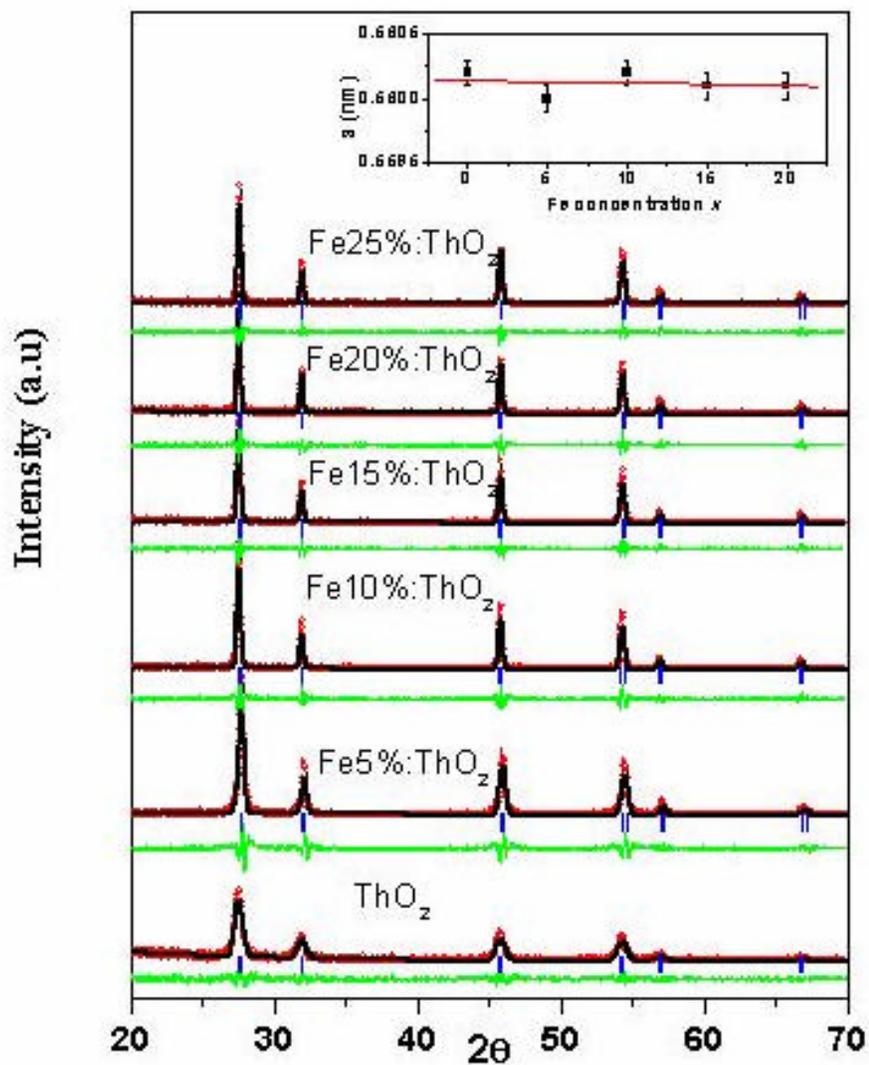

Fig.1 Rietveld refinement of room temperature XRD data of $[Th_{1-x}Fe_x]O_{2-\delta}$ (x = 0.0, 0.05, 0.10, 0.15, 0.20 and 0.25). The open circles represent the observed data while the line through circles is the calculated profile. The vertical ticks below the profile are the expected reflections for fluorite type cubic structure (Fm3m). The continuous line below the vertical tics is the difference pattern. (Inset shows the variation of lattice parameter *a* with dopant concentration *x* up to 20 at. %).

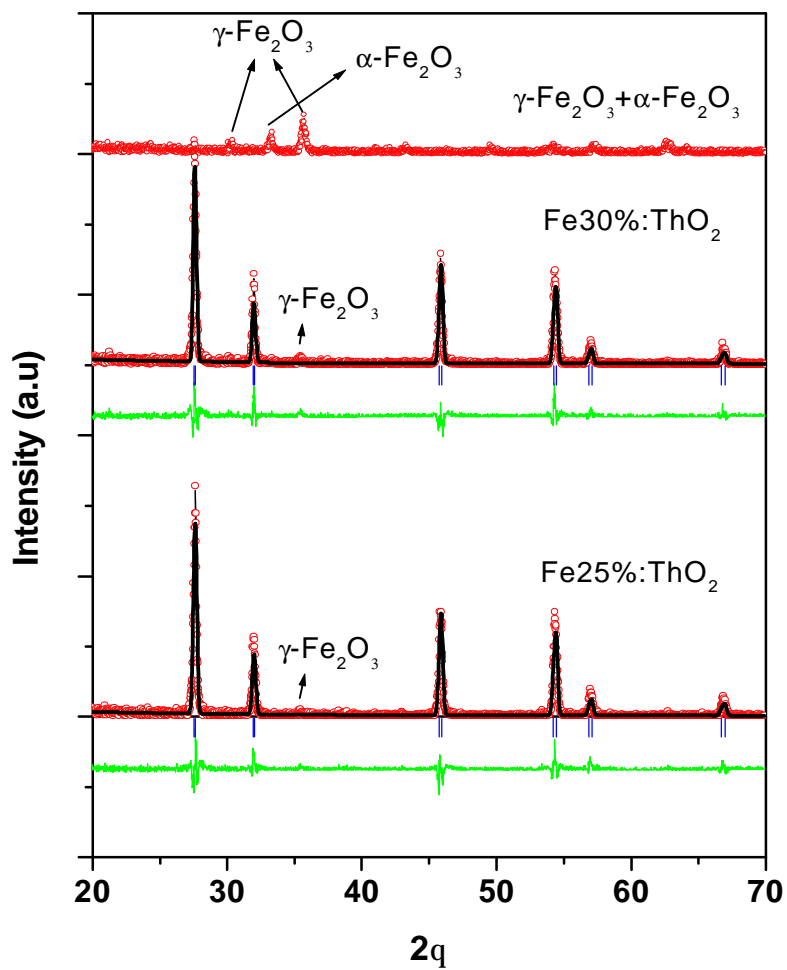

Fig.2 XRD patterns of 25 and 30 at. % Fe doped $ThO_2$ along with XRD patterns of $Fe_2O_3$ (? + α).

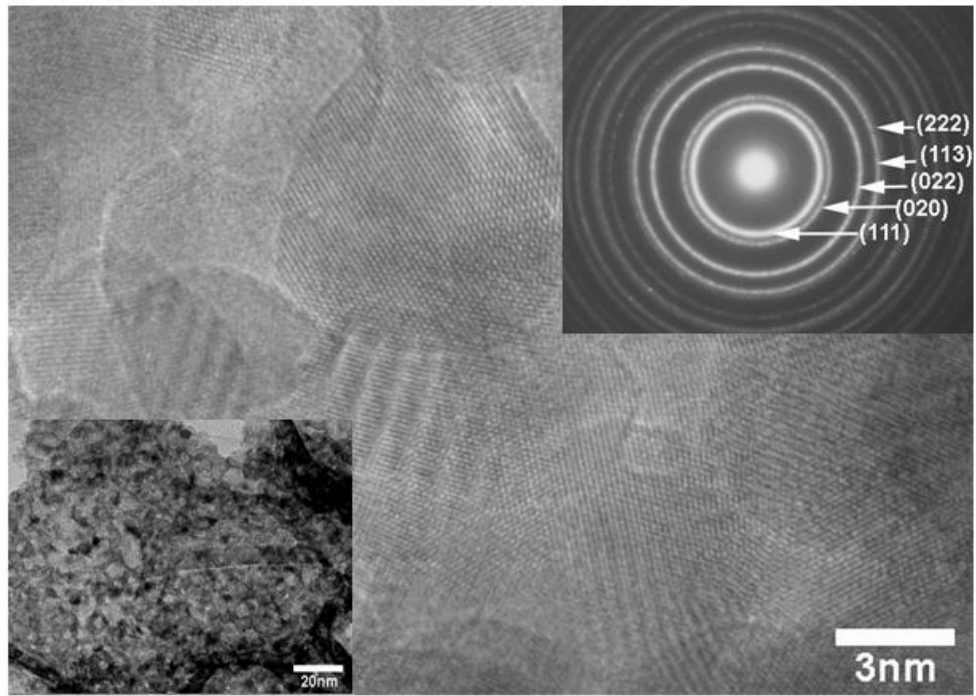

Fig.3   HRTEM, TEM (inset bottom) and SAED (inset top)of 10 at. % Fe doped ThO$_2$

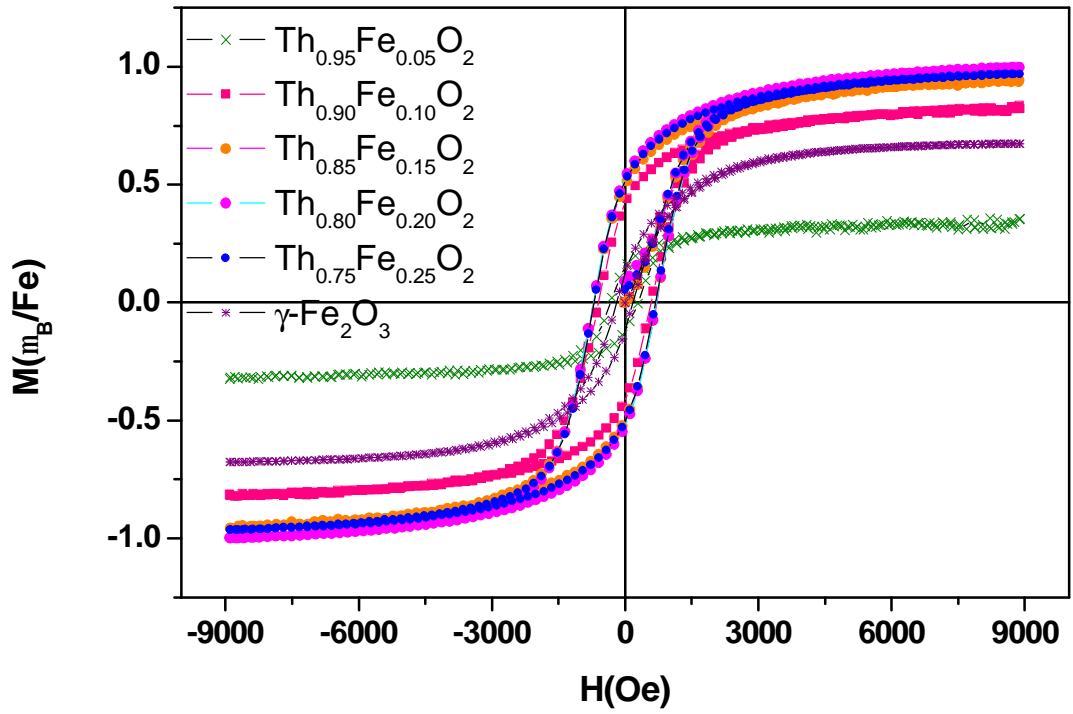

Fig.4  *M* *vs H* curves at RT for $[Th_{1-x}Fe_x]O_{2-\delta}$ (x = 0.0, 0.05, 0.10, 0.15, 0.20 and 0.25) and ?-$Fe_2O_3$.

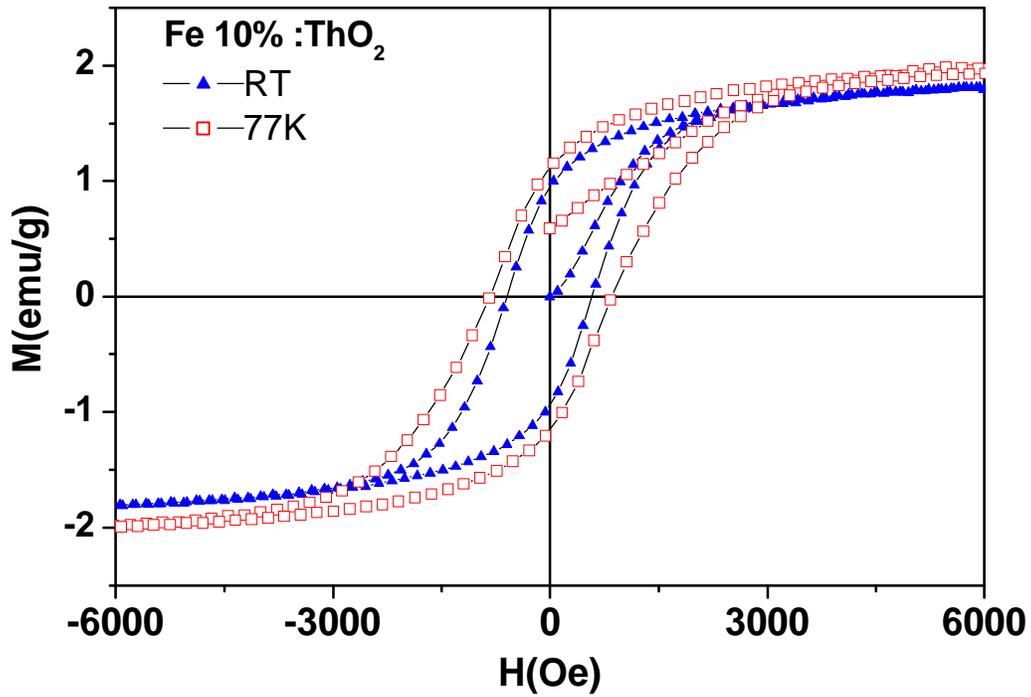

Fig.5  *M vs H* curve at RT and 77 K for 10 at. % Fe doped ThO$_2$.